# Optimal Scheduling of an Isolated Microgrid with Battery Storage Considering Load and Renewable Generation Uncertainties

Yang Li, *Member, IEEE*, Zhen Yang, Guoqing Li, Dongbo Zhao, *Senior Member, IEEE*, and Wei Tian, *Senior Member, IEEE*

***Abstract*—** By modeling the uncertainty of spinning reserves provided by energy storage with probabilistic constraints, a new optimal scheduling mode is proposed for minimizing the operating costs of an isolated microgrid (MG) by using chance-constrained programming. The model is transformed into a readily solvable mixed integer linear programming (MILP) formulation in GAMS via a proposed discretized step transformation (DST) approach and finally solved by applying the CPLEX solver. By properly setting the confidence levels of the spinning reserve probability constraints, the MG operation can be achieved a trade-off between reliability and economy. The test results on the modified ORNL DECC lab MG test system reveal that the proposal significantly exceeds the commonly used hybrid intelligent algorithm with much better and more stable optimization results and significantly reduced calculation times.

## NOMENCLATURE

**Acronyms**

| | |
|---|---|
| MGs | Microgrids |
| ESS | Energy storage system |
| DERs | Distributed energy resources |
| DGs | Distributed generations |
| BESS | Battery ESSs |
| HIA | Hybrid intelligent algorithm |
| CCP | Chance constrained programming |
| MPC | Model Predictive Control |
| MILP | Mixed integer linear programming |
| SOs | Sequence operations |
| PDF | Probability density function |
| ATC | Addition-type-convolution |
| STC | Subtraction-type-convolution |
| ORNL | Oak Ridge National Laboratory |
| DECC | Distributed energy control and communication |
| PSO | Particle swarm optimization |
| MCS | Monte Carlo simulations |
| DST | Discretized step transformation |

**Symbols**

| | |
|---|---|
| $q$ | Discrete step size (kW) |
| $F_C$ | Fuel cost ($) |
| $t$ | A scheduling time period (h) |
| $T$ | Total number of time periods in a cycle (h) |
| $M_G$ | Total number of MT units |
| $C$ | Capacity (kWh) |
| $\eta_{ch}$ | Charge coefficient (p.u) |
| $\eta_{dc}$ | Discharge coefficient (p.u) |
| $\alpha$ | Pre-given confidence level (%) |
| $\tau$ | A large positive number |
| $\varepsilon$ | A small positive number |
| $NO$ | Numbers of probabilistic constraints |
| $NL$ | Numbers of deterministic constraints |

**Subscripts**

| | |
|---|---|
| $w$ | Wind |
| $in$ | Cut-in |
| $out$ | Cut-out |
| $*$ | Rated |
| $r$ | Actual light intensities |
| $max$ | Maximum value |
| $min$ | Minimum value |
| $m$ | Maximum power |
| $pv$ | Photovoltaic |
| $a$ | Probabilistic sequences of WT power outputs |
| $b$ | Probabilistic sequences of PV power outputs |
| $c$ | Probabilistic sequences of joint power outputs of PV and WT |
| $d$ | Load probabilistic sequences |
| $e$ | Equivalent load probabilistic sequences |
| $n$ | Number of MT |
| $T_{end}$ | The end of the entire scheduling cycle |

Manuscript received October 18, 2017; revised March 9, 2018; accepted May 10, 2018. This work was supported in part by the U.S. Department of Energy (DOE)'s Office of Electricity Delivery and Energy Reliability – Advanced Grid Modeling (AGM) Program, the China Scholarship Council (CSC) under Grant 201608220144, the National Natural Science Foundation of China under Grant 51677022 and the National Key Research and Development Program of China under Grant 2017YFB0903400. (Corresponding author: Yang Li.)

Y. Li, Z. Yang and G. Li are with the School of Electrical Engineering, Northeast Electric Power University, Jilin, China. (e-mail: liyang@neepu.edu.cn, 1678084931@qq.com, lgq@neepu.edu.cn ).
D. Zhao is with Energy Systems Division, Argonne National Laboratory, Lemont, USA. (e-mail: dongbo.zhao@anl.gov).
W. Tian is with the Galvin Center for Electricity Innovation, Illinois Institute of Technology, Chicago, IL 60616 USA. (e-mail: wtian3@iit.edu)

| | |
|---|---|
| 0 | Initial energy |
| Ress | Reserve capacity |
| L | Load |

**Superscripts**

| | |
|---|---|
| CH | Charge |
| DC | Discharge |
| CNLOAD | Controllable load |
| WT | Wind turbine |
| PV | Photovoltaic |
| MT | Microturbine |
| EL | Equivalent load |

## I. INTRODUCTION

UNDER the dual pressures of energy demand and environmental protection, renewable energy is experiencing a rapid growth, and a large number of microgrids (MGs) have been invested to power systems during the past fifteen years [1], [2]. An MG is a low- or medium-voltage localized entity consisting of electricity sources, energy storage system (ESS), and loads that operates either in grid-connected or in stand-alone mode [2]. These systems have proven to be promising measures for supplying power electricity to isolated rural villages that are inaccessible to the main power grid and providing critical community services during extreme weather-related incidents involving thunderstorms, hurricanes, and blizzards [3]. By this means, it paves the way for integrating various distributed energy resources (DERs), especially wind turbine (WT), photovoltaic (PV) and microturbine (MT) generators. Therefore, MGs have no doubt become a preferable solution to the energy crises as well as an indispensable complement for the construction of a smart energy city [4].

### A. Literature Review

A significant amount of studies have been carried out on different aspects of MGs, such as planning operation and management, protection and control strategies [5]-[7]. In this study, we focus on the optimal day-ahead scheduling problem for isolated MGs. Due to relatively small capacity and inherently intermittent nature of distributed generations (DGs), the operation of MGs is vulnerable to uncertain power exchange between the sources and the loads, and thereby the operation reliability and supply security are difficult to be guaranteed [8]. In addition, uncertainties from the load side tend to rise with the high-penetration of new power-electronic loads such as electric vehicles [9]. It is necessary to deal with these key challenges in MG optimal scheduling.

To resolve this problem, a number of methods, such as model predictive control (MPC) [10]-[12], robust programming [13]-[15] and stochastic programming [16]-[18], have been proposed in recent years. The MPC is a powerful approach for indirectly handling uncertainty in the load and generation fluctuations, and its main idea is to obtain a control signal via addressing an optimization problem at each time instant with the use of system model [11]. The main disadvantage of such kind of method is the formulation complexity and the fact that an optimization solver package is necessary [10]. Robust programming is a promising method to analyze the worst-case scenario of several uncertainties, and the solution of this approach is guaranteed to be feasible and optimal for a defined uncertainty set [14]. However, the obtained solution is often conservative since it is a hedge against the worst-case realization [15]. Stochastic programming is another effective mathematical tool for coping with the uncertainty that is characterized by a probability distribution on the parameters [16]. This approach requires the probability distributions governing the data are known or can be estimated, but the distributions might be unavailable in practical applications.

More recently, chance-constrained programming (CCP) has gained a great deal of attention for addressing uncertainties in MG operations since it is proven to handle uncertainty more properly especially with probabilistic constraints. In literature, the existing solution methods based on CCP for MG scheduling can be largely divided into two classes: an approximation method [19]-[22], also called hybrid intelligent algorithm (HIA), and an analytical method [11], [23]. The main idea of HIA is to perform large numbers of random samplings by using Monte Carlo simulations (MCSs) to verify the chance constraints and combine the intelligent optimization algorithms to solve CCP problems. In [19], a CCP framework is firstly put forward to model stochasticity of renewable generation and load in the microgrid economic dispatch area, in which the opportunity constraint is used to describe the objective function. Reference [20] proposes a CCP-based probabilistic optimization framework for achieving bi-objective optimal energy management in MGs. In [21], an optimal scheduling strategy for MG economic operation is developed with consideration of the chance-constrained islanding capability. In [22], A CCP-based online optimal control strategy for power flow management in MGs is proposed, and MCSs are utilized for handling the chance constraints. The commonly used HIA suffer from inherent limitations, such as low efficiency and poor stability, which may hinder its practical applications.

The analytical method is another powerful solution tool for solving CCP, and its key idea is to transform a CCP problem into the deterministic equivalent by using the cumulative distribution function (CDF) as well as its inverse function of a random variable and then solve the equivalent deterministic model. In [11], the deterministic equivalent of the chance constraints is built via historical data. In [23], energy management problems under uncertainties for grid-connected MGs are addressed by transforming the CCP model into a deterministic linear program. However, there exist the following open problems in the process of equivalent transformation: first, the CDF and its inverse function are difficult to obtain in many practical cases; second, if there are multiple random variables that obey different distributions in chance constraints, the structure of joint probabilistic sequences distribution of multidimensional random variables are more complicated, which will undoubtedly lead to an increase in the difficulty of equivalent transformation.

In the meantime, research findings have also demonstrated that the ESS plays a significant role in enhancing the malleability, reliability, and resiliency of power systems by serving as a buffer to match supply and demand for integrating DERs [24], [25]. There are many storage technologies, such as battery ESSs (BESS) [24], supercapacitors, fuel cells, and hybrid energy storage systems [25], available to be deployed in MGs. Modern BESS techniques have proved to be suitable for providing ancillary services like rapid spinning reserves to an isolated power system with mature technology and low capital cost [26], [27].

With these reported peer works in the existing literature, fundamental MG optimal scheduling problems are formulated and solved. However, some research gaps still exist in the area as follows. (1) Regarding scheduling models, many of existing investigations focus on utilizing dispatchable generators like MT units to provide spinning reserve services, but not much attention has been paid so far to the role of ESS for doing this for isolated MGs. (2) For solution methodologies, the CDF and its inverse function of a random variable may be unavailable when transforming a chance constraint into the deterministic equivalent. (3) To maintain the system reliability, previous works tend to treat the spinning reserve requirements as deterministic constraints related to the required minimum reserve capacity [28], [29], but doing so will in fact result in an unnecessarily elevated reserve cost if accounting all the uncertainties such as power fluctuations of renewable DGs, load fluctuations and unexpected unit failures or outages [3].

### B. Contribution of This Paper

To address the above concerns, a CCP-based scheduling model and its solution approach are proposed for isolated MGs. The main contributions are as follows:
- The proposed model manages to make full use of ESS to provide spinning reserve service for isolated MGs, and model the uncertainty of spinning reserves with probabilistic constraints related to a risk of constraint violation, rather than traditional deterministic constraints.
- A new discretized step transformation (DST) method is proposed for handling chance constraints: based on discretized probabilistic sequences and their joint probabilistic sequences of random variables via sequence operations (SOs), chance constraints can be directly transformed into their deterministic equivalents without the need of the inverse function of their CDFs, which is a novel methodology for solving CCP model.
- The MG operation can achieve a balanced trade-off between reliability and economy by setting the confidence levels of the spinning reserve probability constraints properly.
- The approach is superior to the commonly used HIA with more stable optimization results and significantly reduced computation, which will be shown in the case study section.

### C. Organization of This Paper

The rest of this paper is organized as follows. An introduction of uncertainty modeling of MGs is given in Section II, followed by serialization modeling of random variables in Section III. The proposed MG scheduling model is shown with an explicit formulation in Section IV, with its solution methodology in Section V. Numerical results are displayed in Section VI, and conclusions are made in Section VII.

## II. UNCERTAINTY MODELING OF MICROGRIDS

### A. Probabilistic WT Model

The uncertainty of WT output is mainly originated from the inherent intermittency of wind speeds. Previous research demonstrates that wind speeds follow the Weibull distribution [26], [29]. The probability density function (PDF) of wind speeds is accordingly given by [24]

$$f_w(v) = (k/\gamma)(v/\gamma)^{k-1} \exp[-(v/\gamma)^k] \quad (1)$$

where $v$ represents the actual wind speed; $k$ is the shape factor (dimensionless), which describe the PDF shape of wind speeds; $\gamma$ is the scale factor.

The relationship between the WT power output $P^{WT}$ and the actual wind speed $v$ can be described as [29]:

$$P^{WT}(v) = \begin{cases} 0 & v < v_{in}, v > v_{out} \\ \dfrac{v - v_{in}}{v_* - v_{in}} P_* & v_{in} \leq v < v_* \\ P_* & v_* \leq v < v_{out} \end{cases} \quad (2)$$

where $P_*$ denotes the rated output power of a WT, $v_{in}$ is the cut-in wind speed, $v_{out}$ is the cut-out wind speed, and $v_*$ is the rated wind speed.

According to (1) and (2), the PDF of the WT output $f_o(P^{WT})$ can be formulated as

$$f_o(P^{WT}) = \begin{cases} (khv_{in}/\gamma P_*)\left[((1+hP^{WT}/P_*)v_{in})/\gamma\right]^{k-1} \times \\ \quad \exp\left\{-\left[((1+hP^{WT}/P_*)v_{in})/\gamma\right]^k\right\}, p^{WT} \in [0, P_*] \\ 0, \quad \text{otherwise} \end{cases} \quad (3)$$

where $h = (v_*/v_{in}) - 1$.

### B. Probabilistic PV Model

The PV output is mainly dependent on the amount of solar irradiance reaching the ground, ambient temperature and characteristics of the PV module itself. The statistical study shows that the solar irradiance for each hour of the day follows the Beta distribution [8], which is a set of continuous probability distribution functions defined in the interval (0, 1). The Beta PDF used to depict the probabilistic nature of solar irradiance is

$$f_r(r) = \frac{\Gamma(\lambda_1) + \Gamma(\lambda_2)}{\Gamma(\lambda_1)\Gamma(\lambda_2)} \left(\frac{r}{r_{max}}\right)^{\lambda_1 - 1} \left(1 - \frac{r}{r_{max}}\right)^{\lambda_2 - 1} \quad (4)$$

where $r$ and $r_{max}$ are respectively the actual and maximum light intensities; $\lambda_1$ and $\lambda_2$ are the shape factors,

$$\lambda_1 = \mu_{pv}\left(\frac{\mu_{pv}(1-\mu_{pv})}{\sigma_{pv}^2} - 1\right), \quad \lambda_2 = (1-\mu_{pv})\left(\frac{\mu_{pv}(1-\mu_{pv})}{\sigma_{pv}^2} - 1\right);$$

$\Gamma$ represents a Gamma function in the following form: $\Gamma(\lambda) = \int_0^{+\infty} \rho^{\lambda-1} e^{-\rho} d\rho$, wherein $\rho$ is an integration variable. The relationship between PV outputs and solar irradiances is [26]

$$P^{PV} = \xi \eta_m A_{pv} \eta_{pv} \cos\theta \quad (5)$$

where $\xi$ is the solar irradiance, $\eta_m$ is the maximum power point tracking, $A_{pv}$ is the radiation area of this PV, $\eta_{pv}$ is the conversion efficiency, and $\theta$ is the solar incident angle.

From (5) it can be seen that the PV output is linear with the solar irradiance, and thereby, the PV output is also generally subject to the Beta distribution. The PDF of PV output is [26].

$$f_p(P^{PV}) = \frac{\Gamma(\lambda_1)+\Gamma(\lambda_2)}{\Gamma(\lambda_1)\Gamma(\lambda_2)}\left(\frac{P^{PV}}{P^{PV}_{max}}\right)^{\lambda_1-1}\left(1-\frac{P^{PV}}{P^{PV}_{max}}\right)^{\lambda_2-1} \quad (6)$$

where $P^{PV}$ and $P^{PV}_{max}$ represent the output of this PV and its maximum value, respectively.

### C. Probabilistic Load Injection Model

A widely used normal distribution model is adopted here for modeling load fluctuations. The PDF can be described as [29]

$$f_l(P^L) = \frac{1}{\sqrt{2\pi}\sigma_L}\exp\left(-\frac{(P^L-\mu_L)^2}{2\sigma_L^2}\right) \quad (7)$$

where $P^L$ is the load active power, $\mu_L$ and $\sigma_L$ are the mean and standard deviation of $P^L$.

### D. Equivalent Load Model

In order to facilitate the incorporation of multiple random variables, the power of an equivalent load (EL) is defined as the difference of the load power and the joint power output of WT and PV, which is expressed as follows:

$$P^{EL} = P^L - (P^{WT} + P^{PV}) \quad (8)$$

where $P^{EL}$ denotes the power of the EL.

## III. SERIALIZATION MODELING OF RANDOM VARIABLES

### A. Introduction of Discretized Step Transformation

The proposed DST is a powerful mathematical tool to handle multiple uncertainties, which is based on the sequence convolution in the field of digital signal processing [30]. The key idea of DST is the concept of sequence operations: first, continuous random variables are discretized as probabilistic sequences according to a given discrete step by using their respective PDFs, and then a newly generated sequence is obtained via mutual operations.

The probabilistic sequences and their mutual operations are described as follows.

**Definition 1.** Suppose a discrete sequence $a(i)$ with the length $N_a$, $a(i)$ is called a probabilistic sequence if

$$\sum_{i=0}^{N_a} a(i) = 1, \ a(i) \geq 0, \ i = 0,1,2,...,N_a \quad (9)$$

**Definition 2.** Given a probabilistic sequence $a(i)$ with the length $N_a$, its expected value is defined as follows:

$$E(a) = \sum_{i=0}^{N_a}[i\ a(i)] = \sum_{i=1}^{N_a}[i\ a(i)] \quad (10)$$

Two kinds of SOs, addition-type-convolution (ATC) and subtraction-type-convolution (STC), are defined as follows.

**Definition 3.** Given two discrete sequences $a(i_a)$ and $b(i_b)$, with length $N_a$ and $N_b$. The ATC and STC are defined as:

$$gs_1(i) = \sum_{i_a+i_b=i} a(i_a)b(i_b), \quad i = 0,1,2,...,N_a+N_b \quad (11)$$

$$gs_2(i) = \begin{cases} \sum_{i_a-i_b=i} a(i_a)b(i_b), & 1 \leq i \leq N_a \\ \sum_{i_a \leq i_b} a(i_a)b(i_b), & i = 0 \end{cases} \quad (12)$$

where $gs_1(i)$ and $gs_2(i)$ are called generated sequences.

### B. Sequence Description of Intermittent DG Outputs

Taking WT as an example, the sequence description of DG outputs are described below. During a time period $t$, the WT output $P_t^{WT}$, PV output $P_t^{PV}$, and load power $P_t^L$ are all random variables, and they can be depicted by the corresponding probabilistic sequences $a(i_{a,t})$, $b(i_{b,t})$ and $d(i_{d,t})$ through discretization of continuous probability distributions. The length of WT output probabilistic sequence $N_{a,t}$ is calculated by

$$N_{a,t} = [P^{WT}_{max,t}/q] \quad (13)$$

where $q$ denotes the discrete step size and $P^{WT}_{max,t}$ is the maximum value of WT power output during period $t$.

Table I shows the WT output and its corresponding probabilistic sequences.

TABLE I  WT OUTPUT AND ITS PROBABILISTIC SEQUENCE

| Power (kW) | 0 | $q$ | … | $u_a q$ | … | $N_{a,t} q$ |
|---|---|---|---|---|---|---|
| Probability | $a(0)$ | $a(1)$ | … | $a(u_a)$ | … | $a(N_{a,t})$ |

The probabilistic sequence of the WT output can be calculated by using its PDF, which is given as follows:

$$a(i_{a,t}) = \begin{cases} \int_0^{q/2} f_o(P^{WT})dP^{WT}, & i_{a,t} = 0 \\ \int_{i_{a,t}q-q/2}^{i_{a,t}q+q/2} f_o(P^{WT})dP^{WT}, & i_{a,t} > 0, i_{a,t} \neq N_{a,t} \\ \int_{i_{a,t}q-q/2}^{i_{a,t}q} f_o(P^{WT})dP^{WT}, & i_{a,t} = N_{a,t} \end{cases} \quad (14)$$

## IV. PROPOSED MG SCHEDULING MODEL

### A. Chance Constrained Programming

As one of the most effective methods for treating uncertainties in optimization, CCP initially proposed by Charnes and Cooper is a formulation of an optimization problem that seeks to achieve the optimal solution under a certain probabilistic constraint [23]. The general formulation of a CCP model can be formulated as

$$\begin{cases} \min \bar{F} \\ s.t. \quad P_{rob}\{F(x,\delta) \leq \bar{F}\} \geq \beta \\ \quad\quad P_{rob}\{G_h(x,\delta) \leq 0\} \geq \alpha \quad h = 1,2...,NO \\ \quad\quad H_j(x) \leq 0 \quad\quad\quad\quad\quad j = 1,2,...,NL \end{cases} \quad (15)$$

where $F(x,\delta)$ is the objective function, $\delta$ is a random variable, $G_h(x,\delta)$ is the uncertainty constraints, $P_{rob}\{\cdot\}$ is the probability of an event occurring, $H_j(x)$ is the traditional deterministic constraints, $\alpha$ and $\beta$ are the pre-given confidence levels, $NO$ and $NL$ are respectively the total numbers of probabilistic and deterministic constraints.

### B. Optimal Scheduling Model

#### 1) Objective Function

The MG scheduling can be modeled as a nonlinear finite-horizon optimal control problem with multiple random variables [11]. Due to the uncontrollability of renewable generations, the MG total operation cost comprises the following three folds: fuel costs of MT units, spinning reserve costs, and charge-discharge costs. Thus, the objective function $F_C$ of the optimal scheduling is modeled as

$$\min F_C = \sum_{t=1}^{T}(g_1(P_t^{DC}) - g_2(P_t^{CH})) + \sum_{t=1}^{T}[\sum_{n=1}^{M_G}(\varsigma_n R_{n,t}^{MT} + \kappa_n S_{n,t} + U_{n,t}(\zeta_n + \psi_n P_{n,t}^{MT}))] \quad (16)$$

where $t$ is a scheduling time period in hours, $P_t^{DC}$ and $g_1(P_t^{DC})$ are respectively the discharge power and cost of the ESS during period $t$, $P_t^{CH}$ and $g_2(P_t^{CH})$ are the charge power and cost, $T$ is the total number of time periods in a scheduling cycle (here $T$ is taken as 24), $n$ is the MT number, $M_G$ is the total number of MT units, $\varsigma_n$ and $\psi_n$ are the MT consumption factors, $U_{n,t}$ and $S_{n,t}$ are 0-1 variables representing respectively the state variable and start-up variable of MT $n$, $\varsigma_n$ and $\kappa_n$ are respectively the spinning reserve cost and the start-up cost of MT $n$, $P_{n,t}^{MT}$ and $R_{n,t}^{MT}$ represent the MT output power and spinning reserve of during period $t$, respectively.

### 2) Constraint Conditions

- *System Power Balance Constraint*

In order to avoid an oversupply of renewable DGs, it is necessary to equip controllable loads for maintaining the power balance in an isolated MG when the DG outputs are relatively large. Consequently, the power balance constraint can be expressed as

$$\sum_{n=1}^{M_G} P_{n,t}^{MT} + P_t^{DC} - P_t^{CH} = E(P_t^{EL}) + P_t^{CNLOAD}, \forall t \quad (17)$$

where $P_t^{CH}$ and $P_t^{DC}$ are the charge and discharge powers of ESS during period $t$, $P_t^{CNLOAD}$ is the output of the controllable load, $P_t^{EL}$ is the predicted value of the EL power, and $E(P_t^{EL})$ is the expected value of $P_t^{EL}$ which is given by

$$E(P_t^{EL}) = \sum_{u_{d,t}=0}^{N_{d,t}} u_{d,t} qa(u_{d,t}) - \sum_{u_{a,t}=0}^{N_{a,t}} u_{a,t} qa(u_{a,t}) - \sum_{u_{b,t}=0}^{N_{b,t}} u_{b,t} qa(u_{b,t}) \quad (18)$$

- *Power Output Constraints of MT units*

The MT output must obey the following inequality:

$$U_{n,t} P_{n,min}^{MT} \leq P_{n,t}^{MT} \leq U_{n,t} P_{n,max}^{MT}, \forall t, n \in M_G \quad (19)$$

where $P_{n,min}^{MT}$ and $P_{n,max}^{MT}$ represent the minimal and maximum outputs of MT unit $n$, respectively.

- *ESS Constraints*

As a widely-used energy storage device for power system applications, a lead-acid battery provides many significant advantages, e.g. a long service life as well as deep charge/discharge cycles [24], and it has been successfully utilized in MGs to balance random fluctuations, maintain system reliability and improve power quality [25-27].

Charge-discharge equation: The relationship between the energy stored in ESS until period $t+1$ and the charging and discharging powers during period $t$ is expressed as [26].

$$C_{t+1} = C_t + (\eta_{ch} P_t^{CH} - P_t^{DC}/\eta_{dc}) \Delta t, \forall t \quad (20)$$

where $C_{t+1}$ and $C_t$ are the energy stored in ESS until period $t+1$ and $t$, $\eta_{ch}$ and $\eta_{dc}$ are respectively the charge-discharge efficiencies. $\Delta t$ is the duration of a time period, which is here taken as 1 hour.

Charge-discharge rate limits: The charge-discharge rate of a lead-acid battery must obey the following constraints:

$$\begin{cases} 0 \leq P_t^{DC} \leq P_{max}^{DC} \\ 0 \leq P_t^{CH} \leq P_{max}^{CH} \end{cases} \forall t \quad (21)$$

where $P_{max}^{CH}$ and $P_{max}^{DC}$ are the maximum charge and discharge power of the ESS during period $t$, respectively.

Capacity limits: The capacity of the lead-acid battery must obey the following constraint:

$$C_{min} \leq C_t \leq C_{max}, \forall t \quad (22)$$

where $C_{max}$ and $C_{min}$ are the maximum and the minimal energy stored in ESS, respectively.

Starting and ending limits [31]:

$$C_0 = C_{T_{end}} = C_* \quad (23)$$

where $C_0$ is the initial energy in ESS, $C_*$ is the initially stored energy limit of ESS, $T_{end}$ is the end of the entire scheduling cycle (it is set to 24h here). For the energy balance of the ESS, the stored energy after the scheduling cycle $C_{T_{end}}$ is set to the same value as the initial stored energy.

- *Spinning Reserve Constraints*

For an isolated MG, due to unavailability of power supports from main grids, the spinning reserve is the most important resource for leveling off the fluctuating power outputs of intermittent DGs and ensuring reliable and economic operation of the system [26], [27]. In this study, besides conventional MT units, the required spinning reserves are also provided by the ESS considering its ability to participate in ancillary services.

Correspondingly, the spinning reserve constraints are

$$P_{n,t}^{MT} + R_{n,t}^{MT} \leq U_{n,t} P_{n,max}^{MT}, \forall t, n \in M_G \quad (24)$$

$$P_{Ress,t} \leq \min\{\eta_{dc}(C_t - C_{min})/\Delta t, P_{max}^{DC} - P_t^{DC}\}, \forall t \quad (25)$$

where $P_{Ress,t}$ is the reserve capacity of the ESS during period $t$.

Equation (17) shows that the overall uncertainties of load and renewable DGs are mainly reflected in $E(P^{EL})$. Hence, to maintain the power balance, the total spinning reserve provided by ESS and MTs is used to compensate for the difference between the fluctuating EL power and its expected value.

In some extreme cases when the joint outputs of intermittent DGs may be zero [3], a sufficient spinning reserve capacity should be provided to maintain system continuity, which will result in a very high reserve cost. However, the probability of such cases occurring is very low. For this reason, it is no doubt a preferable choice to model the spinning reserve requirement as a probability constraint to balance reliability and economy.

$$P_{rob}\left\{\sum_{n=1}^{M_G} R_{n,t}^{MT} + P_{Ress,t} \geq (P_t^L - P_t^{WT} - P_t^{PV}) - E(P_t^{EL})\right\} \geq \alpha, \forall t \quad (26)$$

where $\alpha$ is the pre-given confidence level.

## V. SOLUTION METHODOLOGY

In this section, the methodology is presented which first converts the proposed model into a mixed integer linear programming (MILP) formulation using the DST approach, and then solves it by a CPLEX solver in GAMS.

### A. Model Conversion

#### 1) Probabilistic Sequence of Equivalent Load Power

In order to transform (26) into an equivalent deterministic constraint, the prerequisite is the probability distribution of the random variable $Z=P^L - P^{WT} - P^{PV}$ and its inverse transform [23]. For ease of description, we introduce a variable $X=P^{WT}+P^{PV}$. The probability distribution $F_Z(z)$ is

$$\begin{cases} F_Z(z) = \int_{-\infty}^{z} [\int_0^{P_{max}^L} f_X(P^L - z)f_l(P^L)dP^L]dz \\ f_X(x) = \int_0^{P_{max}^{PV}} f_p(P^{PV})f_o(x - P^{PV})dP^{PV} \end{cases} \quad (27)$$

where $f_p(\cdot)$, $f_l(\cdot)$ and $f_o(\cdot)$ are the PDFs of $P^{PV}$, $P^L$ and $P^{WT}$.

However, the determination of the inverse transform $F_Z^{-1}(z)$ is a critical challenge due to the complex form of the PDFs listed above. In addition, there may exist multiple solutions during the transformation process [11], [23]. To solving such a problem, SOs are introduced to discretize the probability distribution of the random variables.

It is assumed that uncertainties of WT, PV and loads are independent in this work, which is a common hypothesis. Given that the probabilistic sequences of $P_t^{WT}$ and $P_t^{PV}$ are respectively denoted as $a(i_{a,t})$ and $b(i_{b,t})$, the probabilistic sequence $c(i_{c,t})$ of the joint power outputs of PV and WT is obtained by the ATC operation

$$c(i_{c,t}) = a(i_{a,t}) \oplus b(i_{b,t}) = \sum_{i_{a,t}+i_{b,t}=i_{c,t}} a(i_{a,t})b(i_{b,t}), \; i_{c,t} = 0,1,...,N_{a,t}+N_{b,t} \quad (28)$$

Given the probabilistic sequence of $P_t^L$ is $d(i_{d,t})$ with length $N_{d,t}$, the probabilistic sequence of EL power $e(i_{e,t})$ is calculated by the STC operation:

$$e(i_{e,t}) = d(i_{d,t}) \ominus \begin{cases} \sum_{i_{d,t}-i_{c,t}=i_{e,t}} d(i_{d,t})c(i_{c,t}), & 1 \leq i_{e,t} \leq N_{e,t} \\ \sum_{i_{d,t} \leq i_{c,t}} d(i_{d,t})c(i_{c,t}), & i_{e,t} = 0 \end{cases} \quad (29)$$

The corresponding relation between the probabilistic sequence of the equivalent load power $P^{EL}$ with the step size $q$ and the length $N_{e,t}$ is shown in Table II.

TABLE II PROBABILISTIC SEQUENCE OF EL POWER

| Power (kW) | 0 | $q$ | … | $u_e q$ | … | $N_{e,t} q$ |
|---|---|---|---|---|---|---|
| Probability | $e(0)$ | $e(1)$ | … | $e(u_e)$ | … | $e(N_{e,t})$ |

Table II shows that for a given equivalent load power $u_e q$, there is always a corresponding probability $e(u_e)$. All these probabilities constitute a probabilistic sequence $e(i_{e,t})$.

*2) Deterministic Transformation of Chance Constraints*

In order to transform the chance constraint indicated in (26) into deterministic forms, we introduce a new type of 0-1 variable $W_{u_{e,t}}$, which satisfies the following relationship:

$$W_{u_{e,t}} = \begin{cases} 1, & \sum_{n=1}^{M_G} R_{n,t}^{MT} + P_{Ress,t} \geq u_{e,t}q - E(P_t^{EL}) \\ 0, & \text{otherwise} \end{cases} \quad \forall t, u_{e,t} = 0,1,...,N_{e,t} \quad (30)$$

Equation (30) represents that for any period $t$, $W_{u_{e,t}}$ is taken as 1 if and only if the total reserve $\sum_{n=1}^{M_G} R_{n,t}^{MT} + P_{Ress,t}$ is greater than or equal to the difference between the load power $u_{e,t}q$ and its expected value $E(P_t^{EL})$; otherwise, it is taken as 0.

According to Table II, the probability that corresponds to the load power $u_{e,t}q$ is $e(u_{e,t})$. Based on that, (26) is simplified as

$$\sum_{u_{e,t}=0}^{N_{e,t}} W_{u_{e,t}} e(u_{e,t}) \geq \alpha \quad (31)$$

Equation (31) suggests that during period $t$, corresponding to the all the possible output values the EL, the spinning reserve capacity in the MG meets the condition that the required confidence is greater than or equal to $\alpha$. As a result, we can now derive that (31) is equivalent to (26).

The expression of $W_{u_{e,t}}$ in (30) is not compatible with the solution format of MILP. For the sake of facilitating proper resolution of (31), (30) will have to be replaced by the following inequality:

$$(\sum_{n=1}^{M_G} R_{n,t}^{MT} + P_{Ress,t} - u_{e,t}q + E(P_t^{EL}))/\tau \leq W_{u_{e,t}} \leq 1 + (\sum_{n=1}^{M_G} R_{n,t}^{MT} + P_{Ress,t} - u_{e,t}q + E(P_t^{EL}))/\tau, \; \forall t, u_{e,t} = 0,1,...,N_{e,t} \quad (32)$$

where $\tau$ is a large positive number. When the inequality $\sum_{n=1}^{M_G} R_{n,t}^{MT} + P_{Ress,t} \geq u_{e,t}q - E(P_t^{EL})$ holds, (32) is equivalent to the inequality $\varepsilon \leq W_{u_{e,t}} \leq 1+\varepsilon$ (where $\varepsilon$ is a small positive number), while at the same time $W_{u_{e,t}}$ is a 0-1 variable. Therefore, one can drive that the variable $W_{u_{e,t}}$ can only be equal to 1. In a similar manner, if the inequality $\sum_{n=1}^{M_G} R_{n,t}^{MT} + P_{Ress,t} < u_{e,t}q - E(P_t^{EL})$ holds, (32) is equivalent to the inequality $-\varepsilon \leq W_{u_{e,t}} \leq 1-\varepsilon$, and thereby $W_{u_{e,t}}$ can only be equal to 0 for the same reason. In this way, we can deduce that (32) has exactly the same meaning as (30).

Replacing (26) by (31) and (32), the CCP-based model in Section IV has been transformed into a readily MILP formulation.

### B. Solving Process

As displayed in Fig. 1, the solving process of the proposed DST approach mainly includes the follows steps:

Step 1: Model the uncertainties of the MG;

Step 2: Generate the optimal scheduling model using CCP according to (16) ~ (26);

Step 3: Discretize WT and PV outputs and load power as probabilistic sequences;

Step 4: Generate a probabilistic sequence of the equivalent load power via the SOs;

Step 5: Transform the chance constraints into their deterministic equivalents according to (30) ~ (32);

Step 6: Obtain the MILP formulation of the proposed CCP-based model through the above steps;

Step 7: Input the parameters of the MG;

Step 8: Define the transformed model in a problem-solving format in the commercial optimization software GAMS;

Step 9: Solve the model using the branch and bound algorithm provided by the CPLEX solver;

Step 10: Termination criteria. If a solution is found, terminate the process; otherwise, update the confidence level and load, and then return to Step 8.

Step 11: Output the optimal scheduling scheme corresponding to the obtained optimization results.

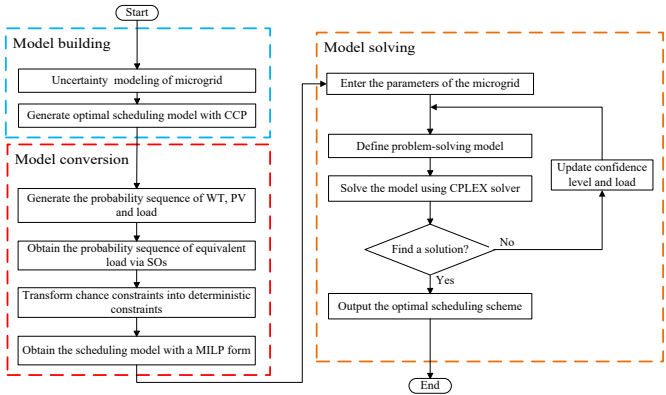

Fig. 1. Flowchart of the proposed solution methodology

## VI. CASE STUDY

The presented approach is tested on a modified Oak Ridge National Laboratory (ORNL) Distributed Energy Control and Communication (DECC) lab microgrid test system [21]. All the simulations are implemented on a PC platform with 2 Intel Core dual-core CPUs (2.4 GHz) and 6 GB RAM.

### A. Introduction of the Test system

As shown in Fig. 2, the system comprises multiple DERs, including a WT unit, a PV panel, three MT units and a battery pack. The parameters of the MT units are listed in Table III.

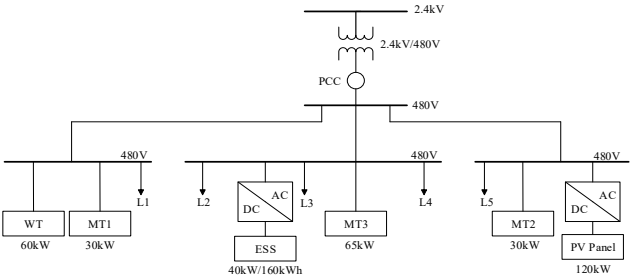

Fig. 2. Modified ORNL DECC microgrid test system

TABLE III PARAMETERS OF MT UNITS

| MT number | $\zeta$ ($) | $\varepsilon$ ($) | $\Psi$ ($/kW) | $\varsigma$ ($/kW) | $P_{min}^{MT}$ (kW) | $P_{max}^{MT}$ (kW) |
|---|---|---|---|---|---|---|
| MT1 | 1.2 | 1.6 | 0.35 | 0.04 | 5 | 30 |
| MT2 | 1.2 | 1.6 | 0.35 | 0.04 | 5 | 30 |
| MT3 | 1.0 | 3.5 | 0.26 | 0.04 | 10 | 65 |

The parameters of the WT are as follows: $v_{in}=3\text{m/s}$, $v_r=15\text{m/s}$, $v_{out}=25\text{m/s}$, $P_r=60\text{kW}$, $P_{min}^{WT}$ and $P_{max}^{WT}$ are the minimal and maximum value of the WT power outputs throughout the entire scheduling period. The parameters of the PV panel are given as follows: $\eta_{pv}=0.093$, $A_{pv}=1300\text{m}^2$, and $P_{max}^{PV}=120\text{kW}$. The parameters of the lead-acid battery are as follows: $P_{max}^{DC}=P_{max}^{CH}=40\text{kW}$, $S_{oc,min}=32\text{kW}\cdot\text{h}$, $S_{oc,max}=160\text{kW}\cdot\text{h}$, $\eta_{dc}=\eta_{ch}=0.9$. The maximum value of the load power is $P_{max}^L=195\text{kW}$. In this study, the charge and discharge prices of ESS are respectively taken as $0.3\$/\text{kW}\cdot\text{h}$ and $0.5\ \$/\text{kW}\cdot\text{h}$.

### B. Analysis and Discussion

#### 1) Operation Costs under Different Confidence Levels

In order to evaluate the change of MG operation costs under different confidence levels, a set of simulations have been repeatedly performed at 11 confidence levels (50%, 55%, ..., 100%) to represent possible changes of confidence levels. The test results are shown in Fig. 3.

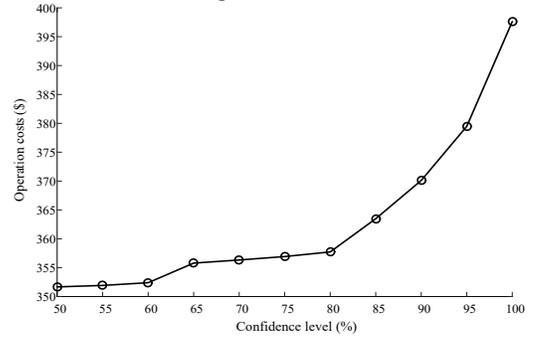

Fig. 3. Operation costs under different confidence levels

As shown in Fig. 3, the MG operation costs increase monotonically from 351.67 $ to 397.61 $ with the increase of the confidence levels from 50% to 100%. Furthermore, the increasing speed of the costs grows faster and faster with an approximately exponential distribution. The reason for this phenomenon is that the confidence level $\alpha$ represents the concept of risk, and there is a trade-off between risk and operation cost. A larger $\alpha$ means reducing the risk of forecast errors of renewable generation and load, while it will inevitably increase the demand for spinning reserves; on the contrary, a smaller $\alpha$ will lead to a low operation cost, but high risks caused by the forecast errors, resulting in the system power imbalance.

The above analysis shows that setting the confidence level $\alpha$ is of practical significance. In fact, due to prediction errors of renewable generations, the MT outputs are unable to strictly meet the load demand and fluctuates at the basic load level. In such cases, the unbalanced powers will have to be balanced by spinning reserves at any time. Therefore, the MG operation achieves a balanced trade-off between reliability and economy by setting a proper confidence.

#### 2) Effect of ESS Charge-discharge Costs

Given the confidence level $\alpha=95\%$ and the load standard deviation $\sigma_L=10\%$, a comparative test with and without consideration of the charge-discharge costs of ESS has been carried out with the results shown in Figs. 4 and 5.

From Figs. 4 and 5, it is revealed that charge-discharge cost of the ESS has an important effect on the MG economic operation. If the charge price is higher than the discharge one, the system absorbs more power from the ESS to participate in the regulation and stabilize the load fluctuations.

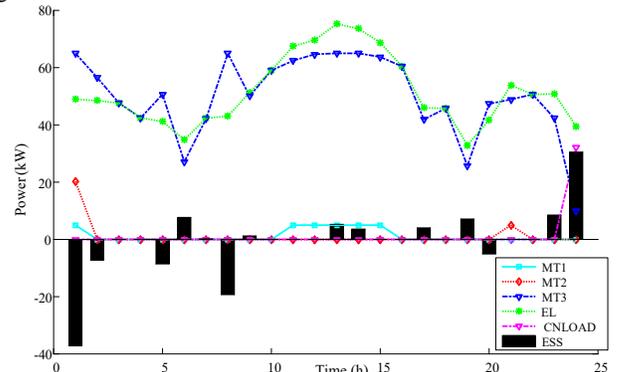

Fig. 4. Scheme with consideration of charge-discharge costs

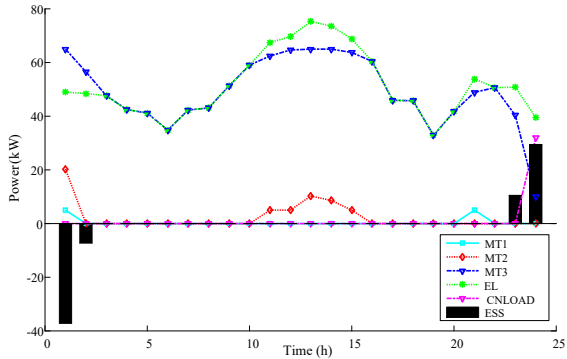

Fig. 5. Scheme without consideration of charge-discharge costs

*3) Spinning Reserves under Different Confidence Levels*

To analyze spinning reserves under different confidence levels, two sets of tests, named Tests I and II, are carried out with load standard deviation $\sigma_L$ =10%. Tests I examine the total spinning reserves, while Tests II further examine the respective spinning reserves provided from MT and ESS.

*Tests I*: The total spinning reserve capacities at different confidence levels are shown in Fig. 6.

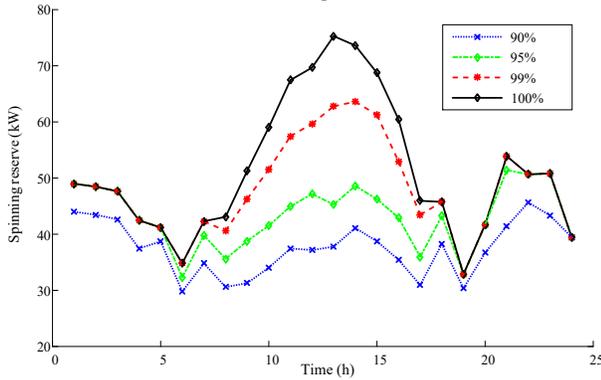

Fig. 6. Spinning reserve capacities under different confidence levels

Fig. 6 indicates that during the joint operation period of the WT and the PV (8-18h), the confidence level can reach 90% when the reserve capacity reaches 35-40 kW. While, during the period of the WT working alone, more spinning reserve capacity needs to be prepared to maintain the 90% confidence level. This demonstrates that the complementary nature of different types of DERs can reduce the fluctuation of output.

With the increased confidence level, the system needs to configure more spinning reserve capacity which inevitably increases the operating costs. Therefore, it is substantial to select the appropriate confidence level to achieve a better balance between reliability and economy.

*Tests II*: The respective spinning reserves provided from MT and ESS at different confidence levels are shown in Fig. 7.

Fig. 7 shows the spinning reserves provided by the ESS are basically higher than those provided by the MTs during different time periods. The reason is that the ESS is given as a priority to provide spinning reserves since it has lower costs and faster response speed than the MTs. Only when the dump energy of the ESS is not affordable for the required spinning reserves, the MTs are utilized to supply reserves.

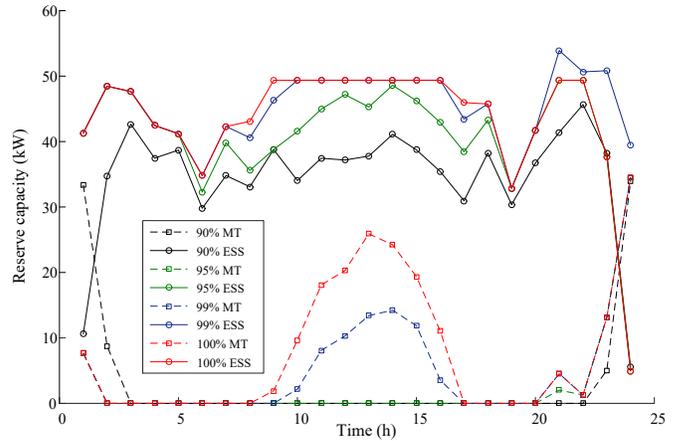

Fig. 7. Reserve capacity from ESS and MG under different confidence levels

From Fig. 7, it can also be seen that with the increase of confidence levels, the both ESS and MTs provides more spinning reserves to balance between demand and supply. For example, when the confidence level rises from 99% to 100%, the spinning reserve provided by the MTs increases dramatically, which will inevitably result in a higher cost.

*4) Influence of ESS Parameters*

To assess the influence of the ESS parameters (i.e., charge-discharge power and energy-storage capacity), two groups of tests are performed with $\sigma_L$ =10%.

The MG operation costs under 25 different charge-discharge powers and energy-storage capacities of the ESS are examined with the results shown in Figs. 8 and 9. Here, the power and capacity are respectively adjusted to the proportions of 30%, 35%, 40%, ..., 150% of the pre-given value of $P_{max}^{DC}$ and $S_{oc,max}$.

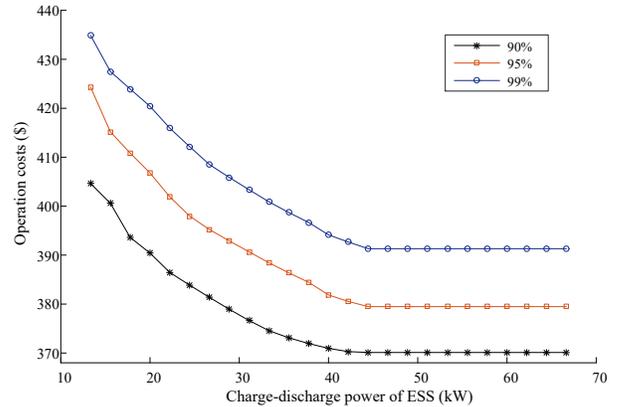

Fig. 8. Operation costs under different charge-discharge powers of the ESS

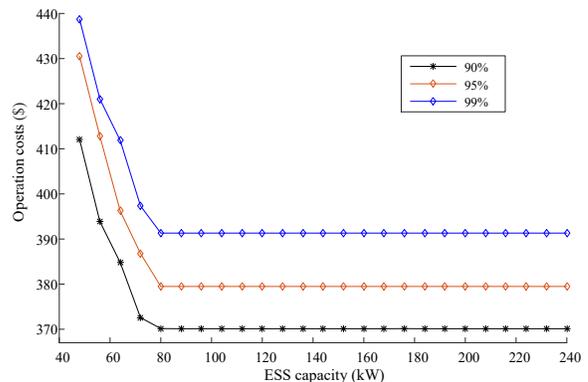

Fig. 9. Operation costs under different energy-storage capacities of the ESS

Figs. 8 and 9 indicate that at the same confidence level, the MG operation cost monotonically decreases with the increase of the ESS charge-discharge power and energy-storage capacities; while the cost gradually increases with the increase of the confidence levels. The results demonstrate that the ESS parameters have a certain influence on the MG operation costs and proper selection of the parameters can reduce the cost.

*5) Impact of Load Fluctuations*

With $\alpha$=95%, the MG operation costs under different load standard deviations $\sigma_L$ from 5% to 20% are shown in Table IV.

TABLE IV  IMPACT OF LOAD FLUCTUATIONS

| $\sigma_L$ (%) | 5 | 10 | 15 | 20 |
|---|---|---|---|---|
| Operation costs ($) | 374.88 | 377.93 | 380.67 | 387.29 |

The results in Table IV suggest that, for a fixed confidence level, the operation costs of the system will correspondingly increase with the increase of load fluctuations. The main reason for this is that the MG system will need more spinning reserves to respond rapidly to variation and keep the power balance between the supply and load slides at any time due to the intensification of load fluctuations [26], which increases the overall operation costs.

*6) Influence of Discrete Steps*

When using the DST, the proper selection of the step size $q$ plays an important role in final optimal results. A smaller step will improve the calculation accuracy, but it will unavoidably introduce the problem of oversized probabilistic sequences, resulting in the dramatic increase in calculation time. On the other hand, a larger step will save calculation time, but it makes the generated sequences unable to fully reflect the actual probability distributions. To this end, an analysis of step-size selection is performed with the results shown in Fig. 10.

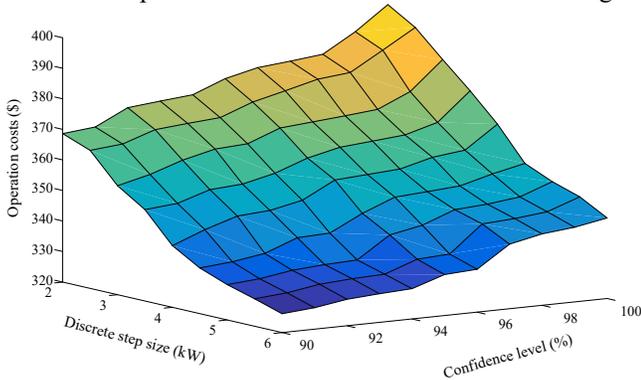

Fig. 10.  Effects of different discrete steps on the optimal results

In Fig.10, when the step size $q$ is greater than 4 kW, the gap between the optimal results is large at the same confidence level. On the contrary, when the step size $q$ is less than 3 kW, the effect of step sizes on optimization results has been drastically reduced, but the computation time sharply increases. Taking into account the above factors, the appropriate range of the step size is bounded by the interval from 3 kW to 4 kW in this study.

*C. Comparison with Hybrid Intelligent Algorithm*

In order to evaluate the performance of the proposed method, the HIA approach which combines the particle swarm optimization (PSO) algorithm with MCS is utilized as a comparison algorithm for solving the model.

The parameters of the HIA are set as follows. For PSO, the population size is 20, and the maximum number of iterations is 150; for MCS, the number of random variables $N_s$ is 500. Due to the uncertainties existing in the results of the HIA [19], [21], the final results of the HIA shown here are the average values of 20 runs.

Let $\alpha$=95%, $\sigma_L$=10%, and the discrete step size $q$ is taken as 2.5 kW. The results are shown in Table V.

TABLE V  COMPARISON RESULTS

| Confidence levels (%) | Proposed approach | | HIA | |
|---|---|---|---|---|
| | Operation cost ($) | Calculation time (s) | Operation cost ($) | Calculation time (s) |
| 90 | 365.2 | 1.9 | 373.7 | 176.4 |
| 95 | 375.4 | 2.1 | 381.2 | 310.3 |
| 100 | 394.3 | 2.0 | 396.5 | 673.5 |

From Table V, the proposed approach outperforms the HIA when solving the proposed MG scheduling model in the following two aspects: first, the operation cost obtained using the proposed approach is lower than that of the HIA; second, the computation time of the proposed method is remarkably less than that of the HIA. Furthermore, with the increasing of the confidence levels, the required computation time of the HIA dramatically roars to derive the optimal solution, while that of the proposed approach remains basically unchanged among the cases studied.

## VII. CONCLUSION

An isolated MG is heavily challenged in its operation and management due to uncertainties in load and renewable generation and the lack of power supports from the main grid. In this paper, we present a new CCP-based scheduling model which make full use of ESS to provide spinning reserve services for isolated MGs. A new DST method is developed for transforming the model into a readily solvable MILP problem, which is then solved by a CPLEX solver in GAMS. Moreover, when handling the spinning reserve requirements, our model utilizes a reserve probability constraint instead of conventional deterministic constraints. This design enables the MG operation to achieve a balanced trade-off between reliability and economy by setting a proper confidence level.

In order to examine the performance of the proposed approach, the modified ORNL DECC lab MG test system is used in the analysis. Test results demonstrate that the approach manages to handle multiple uncertainties, and significantly exceeds the commonly used HIA with more stable solutions and significantly reduced computation.

Future work will focus on the multi-timescale scheduling by integrating day-ahead and real-time scheduling sub-models. Besides, more realistic modeling techniques will be developed to account the correlations between load and DER uncertainties. The coordinated scheduling between MG and an electric vehicle battery charging/swapping station is another interesting topic for future research [32].

ACKNOWLEDGMENT

The authors thank the anonymous reviewers for their careful work and thoughtful suggestions that have helped improve this paper substantially.

**Yang Li** (S'13–M'14) was born in Nanyang, China. He received his Ph.D. degree in Electrical Engineering from North China Electric Power University (NCEPU), Beijing, China, in 2014.
He is an Associate professor at the School of Electrical Engineering, Northeast Electric Power University, Jilin, China. Currently, he is also a China Scholarship Council (CSC)-funded postdoc with Argonne National Laboratory, Lemont, United States. His research interests include power system stability and control, integrated energy system, renewable energy integration, and smart grids.

**Zhen Yang** received his B.S. degree in electrical engineering from the Hebei University of Engineering (HUE), Handan, China, in 2016. Currently, he is pursuing the M.S. degree at the School of Electrical Engineering, Northeast Electric Power University. His research interests include power system stability and renewable energy integration.

**Guoqing Li** received the B.S. and M.S. degrees in Electrical Engineering from Northeast Electric Power University, Jilin, China, in 1984 and 1988, and the Ph.D. in Electrical Engineering from Tianjin University, Tianjin, China, in 1998. Currently, he is a professor at the School of Electrical Engineering, Northeast Electric Power University.

**Dongbo Zhao** (S'10 – M'14 – SM'16) received his B.S. degrees from Tsinghua University, China, M.S. degree from Texas A&M University, College Station, Texas, and the Ph.D. degree from Georgia Institute of Technology, Atlanta, Georgia, all in electrical engineering. Currently, he is an Energy System Scientist with Argonne National Laboratory, Lemont, IL. His research interests include power system control, protection, reliability analysis, and electric market optimization. Dr. Zhao is the Associate Editor of IET Renewable Power Generation, and the Associate Editor of IEEE Access.

**Wei Tian** (M'11-SM'15) received his B.S. degree in Mathematics from Shandong University in 2001, M.S. degree in System Engineering from Xi'an Jiaotong University, China, in 2004, and Ph.D. degree from Illinois Institute of Technology (IIT), Chicago, in 2011. He is a visiting scientist at the Robert W. Galvin Center for Electricity Initiative, IIT. His research interests include power system restructuring, and renewable integration.